\documentclass[sn-mathphys,Numbered]{sn-jnl}
\usepackage[utf8]{inputenc}
\usepackage[T1]{fontenc}

\usepackage[dvipsnames]{xcolor}
\usepackage[commentmarkup=uwave]{changes} 
\definechangesauthor[name=Yan, color=OliveGreen]{YJ}
\definechangesauthor[name=Scott, color=red]{SP}
\definechangesauthor[name=Erwan, color=blue]{EL}

\usepackage{setspace}

\usepackage{graphicx}%
\usepackage{multirow}%
\usepackage{amsmath,amssymb,amsfonts}%
\usepackage{amsthm}%
\usepackage{mathrsfs}%
\usepackage[title]{appendix}%
\usepackage{xcolor}%
\usepackage{textcomp}%
\usepackage{manyfoot}%
\usepackage{booktabs}%
\usepackage{algorithm}%
\usepackage{algorithmicx}%
\usepackage{algpseudocode}%
\usepackage{listings}%
\usepackage{fullpage}
\usepackage{cleveref}
\usepackage{color,soul}
\usepackage{siunitx}

\theoremstyle{thmstyleone}%
\theoremstyle{thmstyletwo}%

\theoremstyle{thmstylethree}%
\usepackage[normalem]{ulem}

\begin{document}

\title[Article Title]{The bandgap-detuned excitation regime in photonic-crystal resonators}



\author[1,2]{\fnm{Yan} \sur{Jin}}\email{yan.jin@colorado.edu}

\author[3]{\fnm{Erwan} \sur{Lucas}}

\author[1,2]{\fnm{Jizhao} \sur{Zang}}

\author[1]{\fnm{Travis} \sur{Briles}}

\author[1,2]{\fnm{Ivan} \sur{Dickson}}

\author[1,4]{\fnm{David} \sur{Carlson}}

\author[1,2]{\fnm{and Scott B.} \sur{Papp}}

\affil[1]{\orgdiv{Time and Frequency Division}, \orgname{National Institute of Standards and Technology}, \orgaddress{\street{325 Broadway MS 688}, \city{Boulder}, \postcode{80305}, \state{Colorado}, \country{USA}}}

\affil[2]{\orgdiv{Department of Physics}, \orgname{University of Colorado}, \orgaddress{\street{390 UCB}, \city{Boulder}, \postcode{80305}, \state{Colorado}, \country{USA}}}

\affil[3]{\orgname{Laboratoire ICB, UMR 6303 CNRS-Université de Bourgogne}, \orgaddress{\postcode{21078} \city{Dijon}, \country{France}}}

\affil[4]{\orgname{Octave Photonics}, \orgaddress{\street{325 W South Boulder Rd Suite B-1}, \city{Louisville}, \postcode{80027}, \state{Colorado}, \country{USA}}}

\abstract{
Control of nonlinear interactions in microresonators enhances access to classical and quantum field states across nearly limitless bandwidth. A recent innovation has been to leverage coherent scattering of the intraresonator pump as a control of group-velocity dispersion and nonlinear frequency shifts, which are precursors for the dynamical evolution of new field states. 
A uniform periodicity nanostructure addresses backscattering with one resonator mode, and pumping that mode enables universal phase-matching for four-wave mixing with control by the bandgap. 
Yet, since nonlinear-resonator phenomena are intrinsically multimode and exhibit complex modelocking, here we demonstrate a new approach to controlling nonlinear interactions by creating bandgap modes completely separate from the pump laser.
We explore this bandgap-detuned excitation regime through generation of benchmark optical parametric oscillators (OPOs) and soliton microcombs. 
Indeed, we show that mode-locked states are phase matched more effectively in the bandgap-detuned regime in which we directly control the modal Kerr shift with the bandgaps without perturbing the pump field.
In particular, bandgap-detuned excitation enables an arbitrary control of backscattering as a versatile tool for mode-locked state engineering.
Our experiments leverage nanophotonic resonators for phase matching of OPOs and solitons, leading to control over threshold power, conversion efficiency, and emission direction that enable application advances in high-capacity signaling and computing, signal generation, and quantum sensing.    
}


\maketitle


Lasers are versatile and precise sources with diverse applications across a range of industries and scientific disciplines, contributing to advanced optical communication~\cite{Jørgensen2022}, atomic and molecular spectroscopy~\cite{Hummon:18}, and quantum technologies~\cite{Kirchmair_2013}.
The capability to generate laser sources with a nearly arbitrary spectrum and quantum-limited noise properties is an enduring challenge that stimulates exploration of new light-matter interactions. Integrated nonlinear photonics plays a key role in enabling the spectrum of light to be dramatically transformed from the input to the output of a device. In particular, nonlinear waveguides transform a mode-locked laser into supercontinuum outputs with exceptionally broad bandwidth at low operating power~\cite{Wu_2024, Jankowski:20, Carlson:2019} and with complex mixtures of nonlinear interactions~\cite{Leefmans_2024}.
Kerr microresonators enable the generation of vastly broadband optical parametric oscillators (OPOs) and soliton microcombs through the conversion of a continuous-wave (CW) pump laser, according to a complex mixture of resonance conditions, group-velocity dispersion of the resonator modes, and the intraresonator pump field~\cite{Briles2018, zang2024laserpower, Black2023, Domeneguetti:21, Cai2022OctavespanningMicrocomb, Rao_2021, Gu:23, Liu_2021, Gong:20}.

Control of nonlinear interactions in microresonators offers the opportunity to engineer new laser sources, especially through foundational behaviors of the intraresonator field~\cite{Lugiato1987}. Group-velocity dispersion (GVD, hereafter dispersion) engineering has been an important advance, enabling direct spectrum control in microcombs~\cite{Lucas2023TailoringMicrocombs, Okawachi2014}. Adjusting the pump laser power and detuning also directly controls a soliton microcomb~\cite{Lucas2017}. Given that solitons are isolated excitations of the microresonator, interacting soliton ensembles offer unique properties~\cite{Cole2016a}. Moreover, ensembles of coupled microresonators enable fundamentally new interactions, including solitons that coexist in several devices~\cite{Yuan2023SolitonPulse}. Direct modification of the intensity-dependent refractive index is also a natural tool for controlling solitons~\cite{Wan_2023, Xu:21}. Among the methods that have been explored to manipulate microcombs, a recent innovation has been photonic-crystal ring resonators (PhCRs), which employ a sub-wavelength nanostructure to couple forward and backward propagation of light and induce mode-frequency shifts~\cite{Yu2020, Lucas2023TailoringMicrocombs}. Indeed, PhCRs provide a rich capability for mode-by-mode dispersion control and to rebalance Kerr frequency shifts.

PhCRs offer direct control of the phase matching and nonlinear dynamics of microresonators. Indeed, access to mode-by-mode frequency shifts represents a fundamentally new degree of freedom for the constituent Lugiato-Lefever equation (LLE) description of a microcomb. In particular, pump-laser excitation of the split mode in PhCRs allows universal phase matching for four-wave mixing in both the normal and anomalous dispersion regimes, enabling OPOs~\cite{Black2022, Liu2023Threshold}, bright solitons in anomalous dispersion~\cite{Yu2020}, access to the dark-to-bright soliton continuum in normal dispersion~\cite{Yu2021}, and use in nonlinear resonator circuits to optimize microcomb performance metrics~\cite{zang2024laserpower}. Recently, PhCR bandgaps for unpumped modes have been used to define the output wavelength in OPOs~\cite{Stone2023WavelengthaccurateNonlinear} and induce exotic microcomb dynamics~\cite{Lucas2023TailoringMicrocombs, Li2023SymmetricallyDispersionengineered}. Although PhCRs intrinsically program mode-by-mode dispersion, since the nanostructured waveguide defines the optical mode, other approaches, including coupled resonators, enable similar controls~\cite{Helgason2023SurpassingNonlinear, Rebolledo-Salgado2023PlaticonDynamics}. Still, the use of coherent backscattering in PhCRs opens excess optical loss channels, particularly since practical devices operate in the large bandgap limit where forward and backward propagation are strongly coupled. These devices include specific limitations, such as increased threshold power, reduced conversion efficiency, and predominant emission of newly generated OPO and soliton light in the backward direction, i.e. emitted toward the pump laser.

Here, we explore the bandgap-detuned excitation regime of PhCRs in which we open optical bandgaps mode-detuned from the pump laser. Operating PhCRs in this regime avoids splitting the pump amongst the forward and backward directions, yet we preserve the capability for universal phase matching. Moreover, by designing bandgaps to interact directly with mode-locked nonlinear states of the resonator, we exploit the inherent multimode composition of target states in their construction. The ratio of forward and backward coupling, and hence the effective excess loss toward the backward direction, is tunable in the bandgap-detuned regime. Through this effect, we introduce a new framework for nonlinear-state engineering based on an effective integrated dispersion parameter, which directly characterizes phase matching of resonator-field states.
We explore benchmark OPOs and soliton microcombs in bandgap-detuned PhCRs, demonstrating that these states are more effectively phase-matched and can be created with reduced threshold power, higher efficiency, and the control to realize predominantly forward emission.
Furthermore, we demonstrate a design process for OPO lasers and solitons microcombs that could be used to create sources for applications.



\begin{figure}[tbp]%
	\centering%
	\includegraphics[width=1\textwidth]{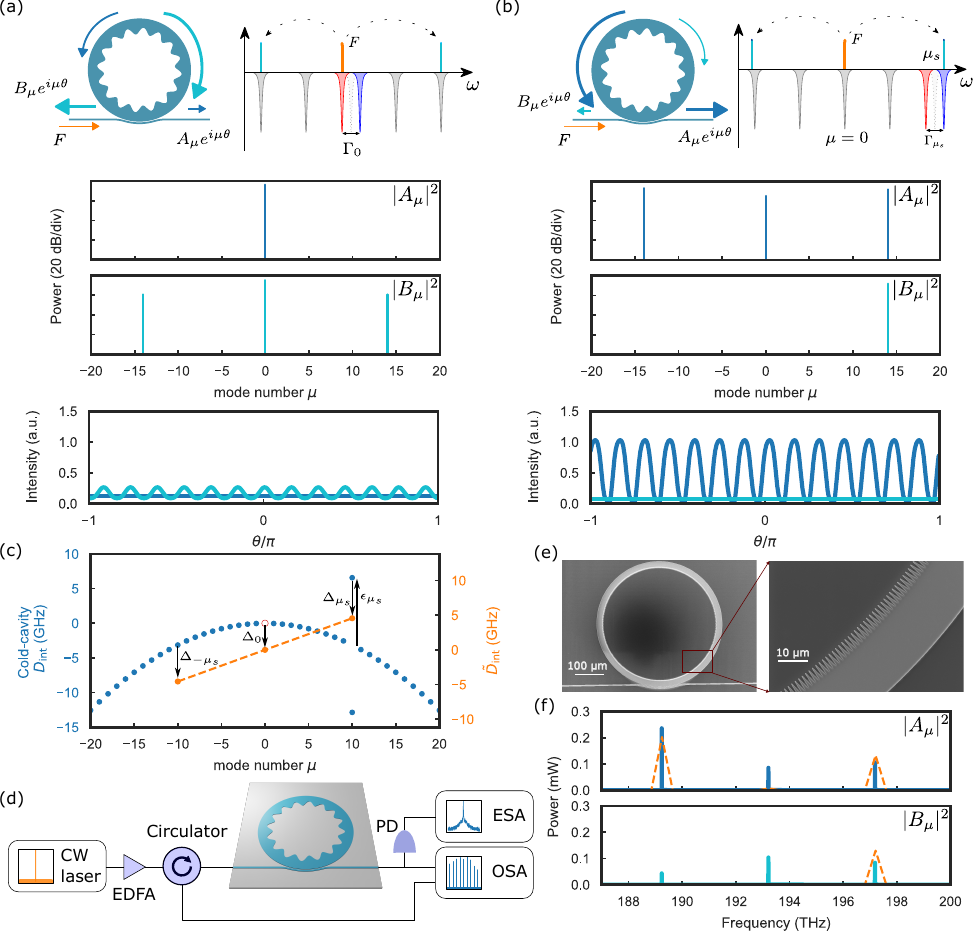}%
	\caption{
		(a) Conventional regime where the red shifted mode is pumped. (b) Bandgap-detuned regime. Both (a) and (b) present the ring structure, $D_{\mathrm{int}}$ normalized to $\kappa/2$, mode structure, the forward and backward spectra, and the pulse intensity normalized to $F^2$. (c) Integrated dispersion for $\mu_s=10$ with low pump power (blue) and high pump power (orange). (d) Setup.
		EDFA: Erbium-doped fiber amplifier.
		OSA: optical spectrum analyzer.
		ESA: electrical spectrum analyzer.
		PD: photodiode.
		(e) SEM images for PhCR. (f) Forward and backward spectra for $\mu_s=10$ in simulation (orange dashed lines) and experiment (solid lines).
	}{}\label{fig1}
\end{figure}

A PhCR is a three-port nonlinear system with an input port receiving the pumping field $F$, and two output ports with modal fields $A_\mu$ and $B_\mu$ propagating respectively in the forward direction (i.e., the same as the pump) and the backward direction, where $\mu$ is the mode number with respect to the pumped mode.
Intraresonator coherent backscattering caused by the periodic sub-wavelength pattern of the PhCR couples the two propagating directions at a specific Bragg criterion and hence at a specific longitudinal mode.
At a designated mode $\mu_s$, this coupling lifts the initial degeneracy of $A_{\mu_s}$ and $B_{\mu_s}$, leading to a red-shifted (lower frequency) and a blue-shifted (higher frequency) resonance, separated by a bandgap $\Gamma_{\mu_s}$.
The dispersion of the resonance frequencies is perturbed and follows the integrated dispersion relation $D_{\mathrm{int}}(\mu)\equiv \omega_\mu - \omega_0 - \mu D_1 = D_2\mu^2/2 \pm \Gamma_{\mu}/2$, where $\omega_\mu$ is the resonance frequencies of the ring resonator, $\omega_0$ is the pump mode frequency, $D_1/2\pi$ is the free spectral range, $D_2$ is the second-order dispersion coefficient, and $\Gamma_\mu=0$ if $\mu\neq\mu_s$.
To describe the dynamics of the PhCR, we introduce the coupled-mode LLEs, written in the spectral domain~\cite{Kondratiev2020ModulationalInstability}:
\begin{align}
	\dot{A}_\mu & = -(1+i\alpha)A_\mu - i\frac{d_2}{2}\mu^2 A_\mu + i\beta_\mu B_\mu + i\mathcal{F} (|A|^2 A)_\mu + i2A_\mu P_b + F\delta_{\mu, 0}
	\label{eq:1}
	\\
	\dot{B}_\mu & = -(1+i\alpha)B_\mu - i\frac{d_2}{2}\mu^2 B_\mu + i\beta_\mu^* A_\mu + i\mathcal{F} (|B |^2B)_\mu + i2B_\mu P_a
	\label{eq:2},
\end{align}
where $F$ refers to the normalized pump field, $\alpha$ is the detuning,  $ \delta_{\mu, 0} $ is the Kronecker symbol, $\kappa$ is the linewidth, $\beta_\mu = \Gamma_\mu/\kappa$ is the normalized half-bandgap in mode $\mu$, $d_2 = 2D_2/\kappa$ is the normalized second-order dispersion coefficient, $P_a = \sum_{\eta}|A_\eta|^2$, $P_b = \sum_{\eta}|B_\eta|^2$, and $\dot{A}_\mu = \mathrm{d}A_\mu/\mathrm{d}\tau$ where $\tau = t\kappa/2$ is the normalized time. $A(\theta, \tau) = \sum A_\mu(\tau)e^{i\mu \theta}$, $B(\theta, \tau) = \sum B_\mu(\tau)e^{i\mu \theta}$ are the normalized electrical fields in the ring coordinates, and $\mathcal{F} (\cdot)_\mu$ represents the Fourier operator.

Our framework for nonlinear-state engineering in PhCRs is based on an effective integrated dispersion $\tilde{D}_\mathrm{int}(\mu)$, which characterizes the frequency mismatch of the field states,
\begin{equation}\label{eq:3}
    \tilde{D}_\mathrm{int}(\mu) \equiv \operatorname{Re}\left(i \frac{dA_\mu/dt}{A_\mu}\right)
    = \frac{\kappa}{2} \operatorname{Re}\left(i \frac{\dot{A}_\mu}{A_\mu}\right) = \alpha \dfrac{\kappa}{2} + \dfrac{D_2}{2} \mu^2 + \epsilon_\mu - \Delta_\mu - P_b\kappa - \delta_{\mu, 0}\operatorname{Im}\left(\dfrac{F}{A_0}\right) \dfrac{\kappa}{2}
\end{equation}
where $\Delta_\mu \equiv \operatorname{Re}(\mathcal{F}(|A|^2A)_\mu/A_\mu)\kappa/2$ is the modal Kerr shift~\cite{Yu2020}, and $\epsilon_\mu \equiv \operatorname{Re}(-\Gamma_\mu \frac{B_\mu}{A_\mu})/2$ is the bandgap-induced frequency shift. The three main contributors in \Cref{eq:3} are the background dispersion $D_2\mu^2/2$, the Kerr nonlinearity $\Delta_\mu$, and the mode shift $\epsilon_\mu$ due to coupling between the couterpropagating fields.
For $F=0$, $\Delta_\mu=0$, $\epsilon_\mu=\pm\Gamma_\mu/2$ indicates two components of the split mode, one blue shifted mode ($\epsilon_\mu>0$) and the other red shifted mode ($\epsilon_\mu<0$) , and $\tilde{D}_\mathrm{int}(\mu)$ reduces to the dispersion of the resonator modes $D_2\mu^2/2 \pm \Gamma_\mu/2$. In mode-locked states, $\tilde{D}_\mathrm{int}(\mu)$ is linear with $\mu$, denoting a stable and stationary field configuration. Therefore, we express $\tilde{D}_\mathrm{int}(\mu) = \mu \omega_\mathrm{rep}$, where $\delta \omega_{\mathrm{rep}}/2\pi = \mathrm{f}_{\mathrm{rep}} - \mathrm{FSR}$ is the difference between the repetition rate of the comb (or line spacing) and the free spectral range. 
In mode-locked states, where $\tilde{D}_\mathrm{int}(\mu)$ and $D_2\mu^2/2$ remain constant, the relative field amplitudes $A_\mu$ and $B_\mu$ are influenced by the interplay between the bidirectional coupling strength $\epsilon_\mu$ and the Kerr nonlinearity $\Delta_\mu$.

In the bandgap-detuned regime, we implement nonlinear state engineering by introducing bandgaps in unpumped modes. First, we consider the case $\Gamma_{\mu}=\Gamma\delta{\mu,\mu_s}$, where $\mu_s\neq 0$ is the split mode. Due to the constraint of $\tilde{D}_\mathrm{int}$ for a mode-locked state, we can control each of the field amplitudes $A_\mu$ or $B_\mu$ with each bandgap $\Gamma_\mu$, which sets the bandgap-induced frequency shift $\epsilon_\mu$. This powerful control parameter also sets the excess power ``lost'' to backscattering in the PhCR. Furthermore, in the bandgap-detuned regime, mode-locked states are excited in either the red mode ($\epsilon_\mu<0$) or the blue mode ($\epsilon_\mu>0$), allowing another new facet to control such states. 
In contrast, in the conventional regime ($\Gamma_\mu = \Gamma \delta_{\mu,0}$) where the red-shifted resonance of the split mode is pumped to generate OPOs or solitons~\cite{Yu2020}, only $\epsilon_0<0$ is allowed.


To understand the nonlinear state excitation in this new scenario, we present benchmark simulations of OPO states in the conventional (\Cref{fig1}(a)) and bandgap-detuned (\Cref{fig1}(b)) regimes. These simulations compare the spectra of forward and backward emission and the intraresonator pulse profiles.
In conventional PhCRs, the generated OPO light propagates in the backward direction ($\sum_{\mu\neq 0}|A_\mu|^2<\sum_{\mu\neq 0}|B_\mu|^2$), while in the bandgap-detuned regime it propagates predominantly in the forward direction.
This difference emerges because at the parametric threshold the backward gain exceeds the forward gain in the conventional regime, while the bandgap-detuned regime is characterized by the dominance of the forward gain.
Furthermore, employing split-mode pumping, as in the conventional regime, results in a notably higher threshold power, nearly twice that of the bandgap-detuned scenario.
By comparing the sideband power between the two regimes in simulation, we find that the bandgap-detuned regime also makes efficient use of the pump energy.

\begin{figure}[h]%
	\centering%
	\includegraphics[width=1\textwidth]{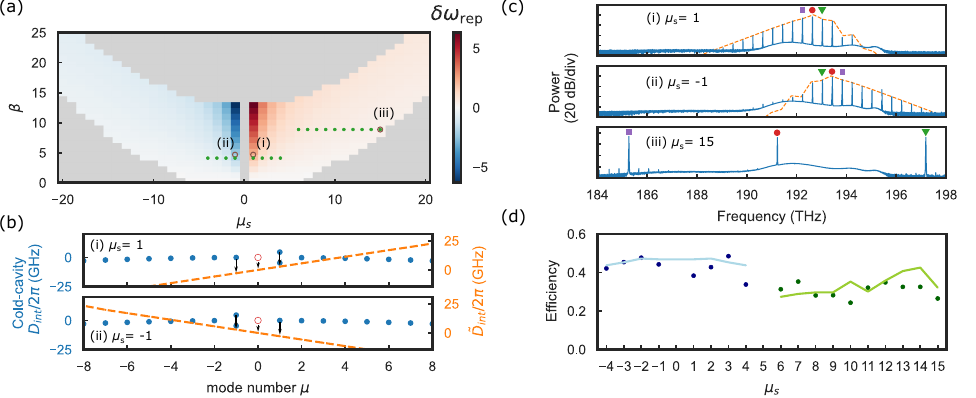}%
	\caption{(a) Existence map with $F=2$, $d_2 = -0.37$.
		The colors indicate the values of $\delta \omega_{\mathrm{rep}}$ as the signal and idler are excited, the brown empty circles marked with (i)(ii)(iii) are explained in (b) and (c), and the green dots refer to the efficiency data points in (d).
		(b) Phase matching diagrams with (i) $\mu_s = 1$ and (ii) $\mu_s =-1$. (c) Simulated (orange) and measured (blue) spectra with (i) $\mu_s = 1$, (ii) $\mu_s =-1$ and (iii) $\mu_s =15$.
		Red circle: pump, green triangle: split mode at $\mu_s$, purple square: mode at $-\mu_s$. (d) Forward efficiency of two devices (blue and green) as we pump different modes.
		The solid lines are the simulated efficiency, and the dots are the efficiency measured in experiments.
	}\label{fig2}
\end{figure}

Since the constraint on $\tilde{D}_\mathrm{int}(\mu)$ is the basis for nonlinear-state engineering, we show in \Cref{fig1}(c) that this quantity explains the formation and advantages of OPOs in the bandgap-detuned regime. 
For $F=0$, we plot the cold cavity effective dispersion $\tilde{D}_\mathrm{int}(\mu) = D_2\mu^2/2 \pm \Gamma_{\mu}/2$  in blue, and the pump mode is indicated by the red circle. Since self-phase modulation is twice smaller than cross-phase modulation in the homogeneous (flat) state, the modal Kerr shift at $0,\pm \mu_s$ obeys $\Delta_0 < \Delta_{\mu_s}, \space \Delta_{-\mu_s}$. Thus, the blue mode at $\mu_s$ is excited so that $\epsilon_{\mu_s} > 0$, and $\tilde{D}_\mathrm{int}(\mu)$ at $\mu=-\mu_s, \space 0, \space \mu_s$ connect directly in a straight-line fashion, as shown by the orange dashed line. This characteristic stands in contrast to the conventional regime, where the red mode of the pumped split mode is excited ($\epsilon_0 < 0$). The split mode $\mu_s$ targets the frequency of the signal or idler, and $\epsilon_{\mu_s}>0$ increases $\Delta_{\mu_s}$. The increased Kerr shift $\Delta_{\mu_s}$ implies that higher power accumulates in the target mode $\mu_s$ and the symmetric mode $-\mu_s$, which explains the enhanced OPOs conversion in the bandgap-detuned regime. The nonzero slope of $\tilde{D}_\mathrm{int}$ indicates a mismatch between the OPO line spacing and the free spectral range, which we will analyze in \Cref{fig2}.

We explore the generation of OPOs and soliton microcombs throughout this paper with the experimental setup in \Cref{fig1}(d).
We generate the pump field with a continuous-wave (CW) laser, amplified by an erbium-doped fiber amplifier (EDFA), and we couple this light into a tantala ($\mathrm{Ta}_2 \mathrm{O}_5$) PhCR~\cite{Jung2019, Black2021}.
An optical circulator separates the forward propagating pump and the backward propagating field from the resonator.
We measure both the forward- and backward-propagating fields with an optical spectrum analyzer (OSA), and we use an electrical spectrum analyzer (ESA) to measure the relative intensity noise of the forward spectra.

We show the physical structure of the PhCR with scanning electron microscope (SEM) images in \Cref{fig1}(e).
The outer sidewall of the ring is completely circular, while we create a spatial modulation on the inner wall of the ring by means of a periodic nanostructure.
We denote the average ring radius as $\mathrm{RR}$, the average width as RW, and the inner corrugated wall is parametrized as $\rho_{\mathrm{in}}(\theta) = \mathrm{RR} - \mathrm{RW}/2 + \rho^{\textsc{phc}} \, \sin(2m\theta)$, where $(\rho, \theta)$ are the polar coordinates, $ \rho^{\textsc{phc}}$ is the amplitude of the photonic crystal and $m$ is the longitudinal order (number of wavelength) of the targeted split mode.
In this paper, $\mathrm{RR}= \qty{54.4}{\micro\meter} $ for all devices so that the FSR is approximately \qty{400}{\giga\hertz}.
$\mathrm{RW}$ varies from \qtyrange[range-units = single]{2}{2.1}{\micro\meter} so that the dispersion coefficient $D_2/2\pi \simeq -10$ MHz, $\rho^{\textsc{phc}}$ is around \qty{5}{\nano\meter}, and $m$ varies to split the modes at different wavelengths.
The ring resonator design shown in this figure is exaggerated so that the SEM images are more clear.

To generate bandgap-detuned OPOs, we tune the pump laser frequency to a mode of the PhCR and monitor the device output. Our framework provides a specific expectation for the resulting state. \Cref{fig1}(f) shows the simulated and experimental spectra for case $\mu_s = 10$, on a linear power scale, highlighting the close correspondence of the traces. 
In the forward direction, both the simulation and the experimental data corroborate that the pump is almost fully depleted, with its power reduced below that of the signal and the idler.
This observation underscores good use of the pump power and the high efficiency of OPO formation in the bandgap-detuned  regime.
In our simulation, only the comb mode at $\mu = \mu_s$ is nonzero in the backward direction, but in the experiment the backward comb lines at $\mu=0, -\mu_s$ also appear due to the reflection of the angled waveguide facets at the edge of the chip.

The central characteristic that controls the generation of bandgap-detuned OPOs is the position of the bandgap relative to the pump $\mu_s$. In particular, we explore phase matching with specific parameters ($F$, $d_2$, $\mu_s$, and normalized half-bandgap $\beta_\mu\equiv \Gamma_\mu/\kappa$) that enable OPO generation.
\Cref{fig2}(a) presents the OPO existence map with respect to the split-mode number $\mu_s$ and the bandgap strength $\beta=\beta_{\mu_s}$ for given driving $F=2$ and dispersion $d_2 = -0.37$.
The gray region indicates the area in which no OPO is generated.
The colored region gives the values of $\delta \omega_{\mathrm{rep}}$ that characterize the difference between $f_{rep}$ and $\mathrm{FSR}$.
Note that this map was not computed via LLE simulations, but by exploring the steady state of the pump mode ($a_0$) and evaluating the dependence of the parametric gain at the modes $A_{\mu_s}$, $A_{-\mu_s}$ and $B_{\mu_s}$ as a function of $\mu_s$ and $\beta$.
The brown circles (marked with (i)(ii)(iii)) and green dots are experimentally measured data points for devices with different $\mu_s$ and $\beta$, which represents our observation points in \Cref{fig2}(b)(c)(d).

In the bandgap-detuned mode-locked regime, the unpumped split modes make the integrated dispersion asymmetric, leading to a non-zero slope coefficient $\delta\omega_\mathrm{rep}$ in $\tilde{D}_\mathrm{int} (\mu)$. The mode detunings and the strength of the bandgaps determine the sign of $\delta\omega_\mathrm{rep}$. 
For OPOs, if $\mu_s>0$, then $\delta \omega_{\mathrm{rep}}>0$ and vice versa.
This is illustrated by the phase matching diagrams in \Cref{fig2}(b); for simplicity, we take (i) $\mu_s = 1$ and (ii) $\mu_s = -1$.
For either case, the blue mode at $\mu = \mu_s$ is excited. 
When the Kerr shift compensates for the original dispersion, if $\mu_s=1$, then $\epsilon_1>0$, and the slope of $\tilde{D}_{\mathrm{int}}(\mu)$ is greater than 0; while if $\mu_s = -1$, then $\epsilon_{-1}>0$ and the slope is negative.
To verify this, we select a device with FSR = \qty{398.078}{\GHz} and pump it into the modes adjacent to the split mode so that $\mu_s$ can be $1$ or $-1$.
The cold cavity dispersion $D_{\mathrm{int}}$ displayed in \Cref{fig2} (b) corresponds to the cases $\mu_s = \pm 1$ for this device, and the first two panels in \Cref{fig2}(c) show the corresponding spectra when we pump these different split modes.
The green triangles point to the split mode $\mu_s$, the red circles indicate the pump mode, and the purple squares indicate the mode at $-\mu_s$.
We then measure the repetition rates (line spacing) and find $f_{\mathrm{rep}}=\qty{398.488}{\GHz}$ for $\mu_s=1$, and $f_{\mathrm{rep}}=\qty{397.569}{\GHz}$ for $\mu_s=-1$.
This corresponds to $\delta \omega_{\mathrm{rep}}/2\pi = \qty{410}{\MHz}$ ($\mu_s = 1$), and $\delta \omega_{\mathrm{rep}}/2\pi = \qty{-509}{\MHz}$ ($\mu_s = -1$).
The difference in the values of $\delta \omega_{\mathrm{rep}}$ is due to different linewidths of the pump mode for $\mu_s = 1$ or $-1$.
In our simulation, we found $\delta \omega_{\mathrm{rep}}/2\pi = \qty{392}{\MHz}$, and \qty{-459}{\MHz} respectively, which is close to our experimental results.

In addition to nonlinear-state engineering, which involves controlling the relative amplitude of $A_\mu$ and $B_\mu$, the bandgap-detuned regime provides the capability to efficiently generate forward-propagating OPOs with varying bandwidth, while minimizing excess loss resulting from backward reflection.
We select a device with a split mode wavelength around \qty{1550}{\nm} and $\mu_s \in \{\pm1, \pm2, \pm3,\pm4\}$, and another device with a split mode around \qty{1521}{\nm} and $\mu_s\in [6,15]$. Their locations on the existence map are marked with green dots in \Cref{fig2}(a). In particular, the data point with $\mu_s=15$ lies at the boundary of the existence map and is marked with a brown circle, showing the prediction ability of the existence map. We plot its forward spectra in panel (iii) of \Cref{fig2}(c). The pump mode is well depleted and has a lower power than the signal and idler, indicating good efficiency performance. \Cref{fig2}(d) shows the efficiency of the two devices in the forward direction as a function of $\mu_s$, obtained by pumping different modes.
The simulations (lines) match the experiments (dots), indicating a high forward efficiency, consistently above 20\%.
In the conventional regime, the best total efficiency can reach 41\% in a complex design that includes a waveguide reflector to recycle the pump field~\cite{Liu2023Threshold}, while in the bandgap-detuned regime, forward efficiency alone can reach 48\%.


\begin{figure}[ht!]%
	\centering%
	\includegraphics[width=1\textwidth]{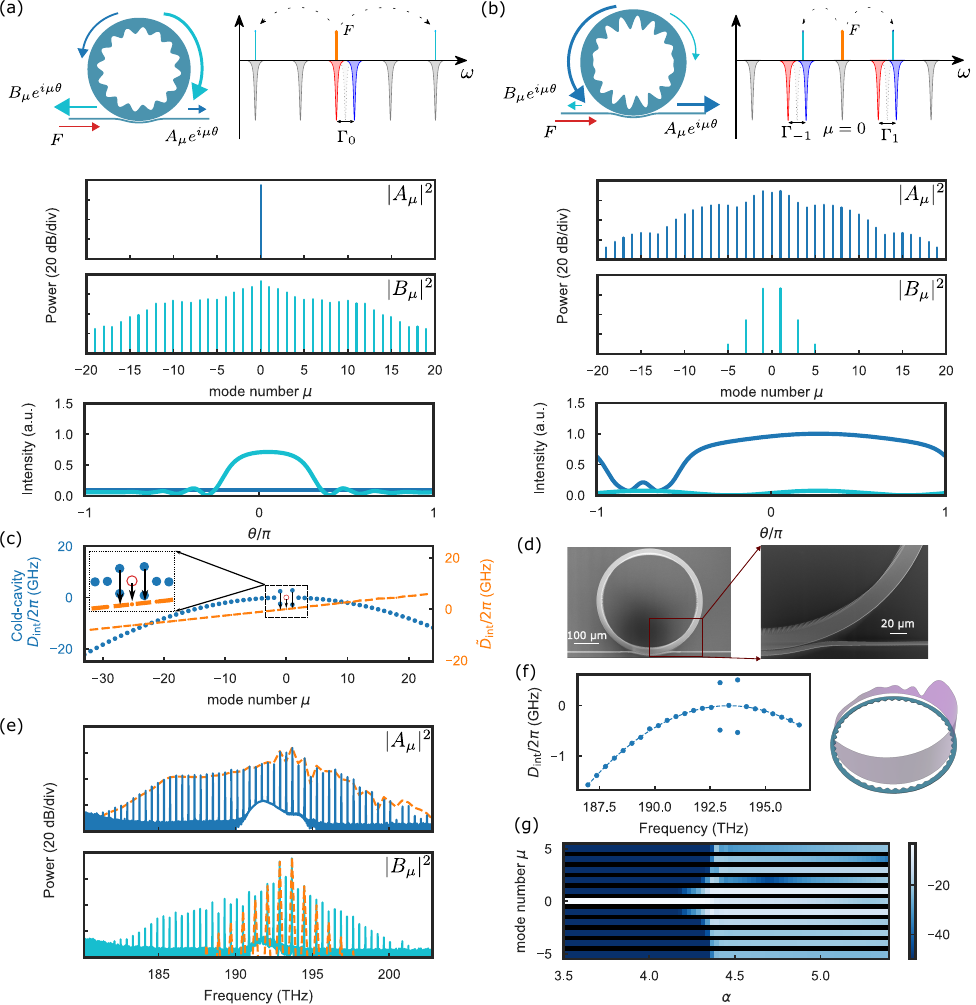}%
	\caption{
		(a) Conventional regime of generating combs in a photonic-crystal ring resonator.
		The red mode of the pump mode is on resonance and the backward combs dominate.
		(b) Bandgap-detuned excitation regime.
		Modes $\mu_s = \pm 1$ are split, and the simulated spectra and pulse indicate the forward-propagating combs. In (a)(b), the blue traces indicate the forward spectra or pulse, and the cyan traces indicate the backward spectra or pulse.
		(c) Phase-matching diagram for the bandgap-detuned regime in the cold and hot cavities. (d) SEM images. (e) Experimental (solid) and simulated (orange dashed) spectra in the forward ($|A_\mu|^2$) and backward  ($|B_\mu|^2$) direction for the bandgap-detuned regime.
		(f) Measured dispersion and simulated pulse on the ring resonator.
		(g)	Simulated spectra with respect to $\alpha$. The unit of the map is dB.
	}\label{fig3}
\end{figure}

We now turn to the generation of soliton microcombs in normal dispersion with the bandgap-detuned scheme. By designing PhCRs with multiple unpumped bandgaps, we open up mode-locked-state engineering of solitons, following the same analysis based on the frequency mismatch operator $\tilde{D}_\mathrm{int}$ of the mode fields in the device.
We begin by comparing soliton generation in PhCRs with conventional (\Cref{fig3}(a)) and bandgap-detuned (\Cref{fig3}(b)) excitation. With the same pump power setting $F$, we calculate the evolution of the forward (blue) and backward (cyan) propagating comb fields as the pump tunes into the mode $\mu=0$, into the lower frequency resonance in the conventional case ($\Gamma_0 > 0$) or to a bandgap-detuned scenario with equal symmetric bandgaps $\Gamma_{\pm1}$ at $\mu_s=\pm1$.
We obtain the comb spectra $|A_\mu|^2$ and $|B_\mu|^2$, and the corresponding intraresonator temporal profiles in the forward (blue) and backward (cyan) directions. In the conventional PhCR scheme, the intracavity field evolves from primary OPO sidebands to a soliton microcomb in the backward direction.
In the case of bandgap-detuned PhCR, parametric gain arises in the split modes $\mu_s$, resulting in the direct generation of sidebands into these modes. This initial modulation further evolves into a deterministic, forward-propagating dark soliton, made possible by the non-split nature of the pumped mode, ensuring that most of its power propagates forward. Conversely, the backward-coupled comb modes at $\mu_s=\pm 1$ give rise to a weaker 2-FSR spaced comb in this direction.

To understand phase matching and the formation of bandgap-detuned soliton microcombs, we plot $\tilde{D}_\mathrm{int}(\mu)$ in \Cref{fig3}(c). Considering an example where the bandgaps at $\mu_s = \pm1$ are not equal ($\Gamma_1>\Gamma_{-1}$) highlights an important practical element of bandgap-detuned microcombs~\cite{Li2023SymmetricallyDispersionengineered}. In this case, $D_\mathrm{int}(\mu)$ is plotted in blue, and the red circle indicates the pump mode. 
At threshold, the modal Kerr shift obeys $\Delta_1\simeq \Delta_{-1}>\Delta_0$ because cross-phase modulation is larger than self-phase modulation, which is shown by the black arrow. As we increase the pump power, the modal Kerr shift $\Delta_\mu$ compensates for the dispersion $D_\mathrm{int}$ with the help of the mode splittings at $\mu_s=\pm 1$,
thus the phase matching is achieved and $\tilde{D}_\mathrm{int}(\mu)$ evolves into a straight line (orange). 
The mode in $\mu = 1$ is more blue-shifted than in $\mu = -1$, therefore, $\epsilon_1>\epsilon_{-1}$, thus $\tilde{D}_{\mathrm{int}}(\mu)$ exhibits a positive slope in the mode-locked state, increasing the repetition rate of the comb ($\delta \omega_{\mathrm{rep}}>0$). 

To experimentally explore the generation of bandgap-detuned dark-soliton, we study the PhCR device in \Cref{fig3}(d) with bandgap modes $\mu_s = \pm1$. We design this resonator by superposing two grating patterns in the inner wall of the ring that are written as a parametric curve~\cite{Lu2020, Lucas2023TailoringMicrocombs}: $\rho_{\mathrm{in}}(\theta) = \mathrm{RR - RW/2} + \sum_{\mu\in\mu_s}\rho^{\textsc{phc}}_\mu \, \sin(2(m_0 + \mu)\theta)$.
We visualize the fabricated pattern by imaging a device with an exaggerated amplitude $\rho$, using a high-resolution SEM. We characterize $D_\mathrm{int}$ of the fabricated device by a calibrated frequency scan of our tunable laser across several resonator modes; see \Cref{fig3}(f). In particular, we resolve the resonator dispersion and the bandgaps $\Gamma_1/2\pi = \qty{1.04}{\GHz} $, $\Gamma_{-1}/2\pi = \qty{0.94}{\GHz} $, $D_1/2\pi = \qty{397.848}{\GHz}$, and $D_2/2\pi = \qty{-12.3}{\MHz}$. On the basis of this information, we simulate the intraresonator intensity pattern of the bandgap-detuned soliton. Experimentally, we pump this PhCR with a power corresponding to $F=2.5$ and slowly tune the pump laser frequency on resonance, starting from the blue-detuned side. This procedure yields a soliton microcomb propagating in the forward direction, which we characterize by measuring the optical spectrum of the forward (solid blue line) and backward (solid teal line) outputs from the device. Comparisons to our simulation (orange line in \Cref{fig1}e) show a good match between the predicted forward comb and the experiment, confirming the close connection between the designed bandgaps, our mode-locked state engineering with $\tilde{D}_{\mathrm{int}}(\mu)$, and the operational generation of solitons in such devices. As we predict from $\tilde{D}_{\mathrm{int}}(\mu)$ in \Cref{fig3}(c), the larger bandgap at $\mu=1$ than $\mu=-1$ leads to $\delta \omega_{\mathrm{rep}}>0$, so there are more modes with positive $\Delta_\mu$ for $\mu<0$ than $\mu>0$, leading to an asymmetric comb with a flatter comb profile for $\mu<0$, as shown in the forward combs in \Cref{fig3}(e). Indeed, we measured $\delta\omega_\mathrm{rep}/2\pi$ to be 21~MHz, consistent with our prediction that $\delta\omega_\mathrm{rep}>0 $ in this device. Regarding the measured spectra and simulation in the backward direction, chip facet reflections obscure the observation of the entire backward spectrum. Nonetheless, characteristic spectral peaks in the modes $\pm1$ are prominently evident. We obtain a conversion efficiency from pump to comb of \qty{37}{\percent} in the forward direction and a \qty{15}{\percent} conversion efficiency in the backward direction, which underscores the fundamental benefits of bandgap-detuned excitation. This fraction of backscattering enables mode-locked-state engineering with the designed bandgaps, creating the forward propagating dark soliton. \Cref{fig3}(g) shows the simulated spectra, focusing on the first few lines, upon scanning the detuning $\alpha$. It verifies our prediction that the comb lines at $\mu_s = \pm 1$ are generated first and then trigger the development of the entire comb, which is different from traditional soliton generation where the primary combs are formed away from the pump.

\begin{figure}[hb!]%
	\centering%
	\includegraphics[width=1\textwidth]{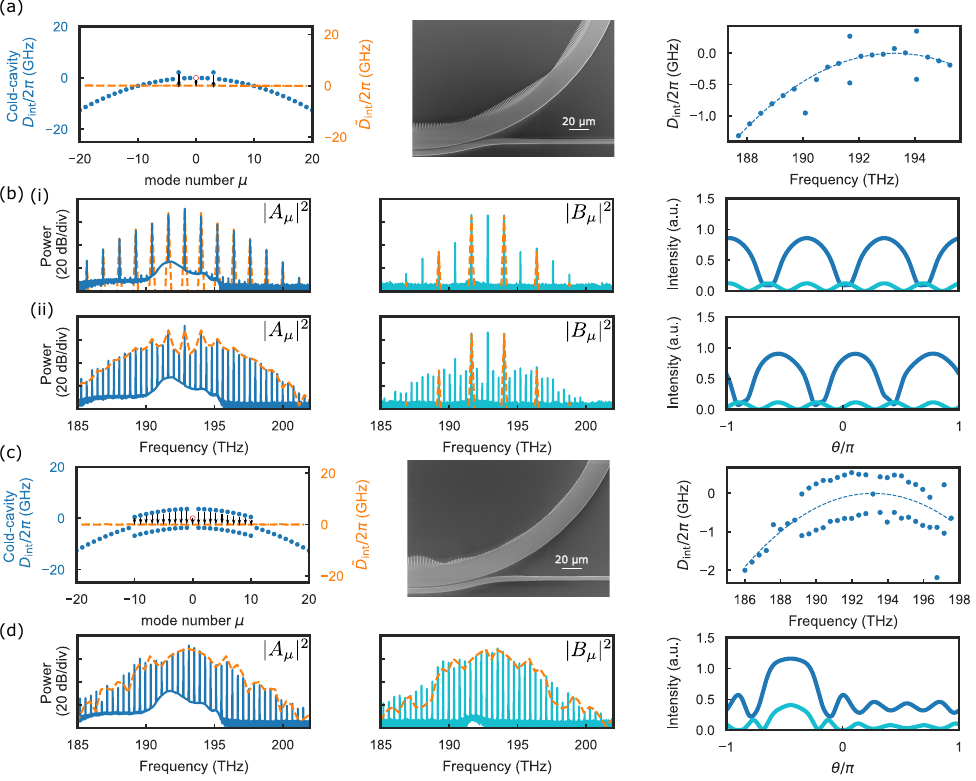}%
	\caption{Photonic-crystal ring resonator with
		(a)  $\mu_s = \pm 3$. The OPOs and soliton crystals generated with this design are shown in (c).
		(b)  $\mu_s = \pm 1, 2, ..., 10$. The bright soliton generated with this design is shown in (d).
		From left to right: Phase matching diagram; pulse shape; SEM images; measured $D_\mathrm{int}$.
		(c) Spectra and pulses for design in (a) with (i) low and (ii) high detuning.
		(d) Spectra and pulses for design in (b).  For (c) and (d), forward ($|A_\mu|^2$) and backward ($|B_\mu|^2$) spectra are plotted in solid lines, and orange dashed lines are simulation results. The forward and backward pulses are plotted in the same color as the spectra in the same direction. }\label{fig4}
\end{figure}

Expanding on this framework for mode-locked-state engineering of dark solitons, we explore other states by way of design with $\tilde{D}_\mathrm{int}$. Important examples include soliton crystals~\cite{Cole2016a} and bright pulses in normal dispersion~\cite{Yu2021}. To create a soliton crystal in the bandgap-detuned regime, we place bandgaps at $\mu_s=\pm 3$. \Cref{fig4}(a) shows the phase-matching diagram with the dispersion of the designed resonator and the simulated $\tilde{D}_\mathrm{int}$ when soliton crystals are generated.
We show an exaggerated PhCR nanostructure with high-resolution SEM, and we characterize $D_{\rm{int}}$ with a calibrated laser frequency scan. To generate soliton crystal microcombs, we pump the device at the level of $F=2.4$ by scanning the laser frequency into the mode $\mu=0$.
\Cref{fig4}(b) shows spectra and soliton-crystal formation at an initial detuning setting (i) and then at a different value of $\alpha$ (ii) where the laser frequency was slightly reduced.
Because phase matching is designed to occur on modes $\mu_s = \pm 3$, the initial modulation pattern (i) is composed of three periods in the resonator, leading to the formation of a soliton crystal with three pulses of equal temporal space in the resonator~\cite{Cole2016a}.
When the laser frequency is decreased, the pulse spacing is perturbed, leading to the appearance of natively spaced lines in the spectra (see panel (ii)), as observed in soliton crystal combs with a defect~\cite{Cole2016a}.
We attribute this phenomenon to the interaction between the oscillating tails of the switching waves that form dark pulses~\cite{Garbin2017}.

We can form bright pulses in the bandgap-detuned regime by splitting many modes of the PhCR except for the pump mode; as shown in \Cref{fig4}(c). These split modes individually control the relative amplitude of $A_\mu$ and $B_\mu$, creating the dispersion profile needed for bright solitons and controlling the excess loss to coherent backscattering.
We study an example in which twenty modes $|\mu_s| \leq 10$ except the pump are split by an equal amount to create a so-called meta-dispersion in the mode spectrum~\cite{Lucas2023TailoringMicrocombs}. Despite the large number of split modes, the principle for phase matching remains the same, and more phase matching pathways are enabled.
At the parametric oscillation threshold, the entire higher frequency branch of the $D_{\rm{int}}$ spectrum is shifted to lower frequency due to the Kerr effect. Therefore, a soliton forms as the pump mode and the blue-shifted branch align, creating a $\tilde{D}_{\mathrm{int}}$ profile that is a straight line. 
The associated photonic crystals form a more intricate corrugated pattern in the ring resonator as shown in the SEM image in \Cref{fig4}(c).
The measured dispersion verifies that the targeted mode structure is implemented correctly and that the pump mode is not split.
Interestingly, this split mode distribution consistently creates a single pulse in the ring resonator.
\Cref{fig4}(d) presents the spectra from simulation and experiments, which allow us to characterize the generated state with respect to the $\tilde{D}_{\mathrm{int}}$ design.
From the spectral data, we find that the forward efficiency is \qty{24}{\percent} and the backward efficiency is \qty{14}{\percent}, resulting from the more equal ratio of forward and backward emission of this design.
Despite an increased fraction of the comb power emitted in the backward direction because multiple modes are split, more power accumulates in the modes $|\mu|\leq 10$ and create a bright pulse within this PhCR.
The comb also features a relatively flat spectral intensity profile and the distribution of several split modes about the pump mitigates imbalance in the bandgaps of fabricated devices, creating a symmetrical soliton microcomb profile.

In summary, we have explored the bandgap-detuned excitation regime of PhCRs, demonstrating modelocked state engineering by design of modal Kerr shifts. The bandgap-detuned regime supports a rich state space of unique OPOs and soliton microcombs that offer low threshold power, high efficiency, and control over forward and backward emission.
In exchange for controlled loss directed backward, the bandgaps on the side modes facilitate effective phase matching in the forward direction, optimizing efficiency. Notably, the forward efficiency of OPOs reaches \qty{48}{\percent}, while that of the dark soliton achieves \qty{37}{\percent}. Although bright solitons exhibit a slightly lower forward efficiency (\qty{24}{\percent}) due to the splitting of more modes, this regime unveils new possibilities for comb states.
The bandgap-detuned excitation regime exhibits significant promise in realizing the application potential of OPOs and microcombs, spanning areas such as spectroscopy, telecommunications, computing, and quantum sensing.



\clearpage
\singlespacing

\section*{Acknowledgements}
We thank Grant Brodnik and Lindell Williams for reviewing the manuscript. This work is funded by the DARPA PIPES program HR0011-19-2-0016, the DARPA QuICC program FA8750-23-C-0539, the DARPA NaPSAC program, AFOSR FA9550-20-1-0004 Project Number 19RT1019, NSF Quantum Leap Challenge Institute Award OMA - 2016244, and NIST. This work is a contribution of NIST and not subject to US copyright. Mention of specific companies or trade
names is for scientific communication only and does not constitute an endorsement by NIST.

\bibliography{ref.bib,library.bib}


\begin{thebibliography}{37}
\ifx \bisbn   \undefined \def \bisbn  #1{ISBN #1}\fi
\ifx \binits  \undefined \def \binits#1{#1}\fi
\ifx \bauthor  \undefined \def \bauthor#1{#1}\fi
\ifx \batitle  \undefined \def \batitle#1{#1}\fi
\ifx \bjtitle  \undefined \def \bjtitle#1{#1}\fi
\ifx \bvolume  \undefined \def \bvolume#1{\textbf{#1}}\fi
\ifx \byear  \undefined \def \byear#1{#1}\fi
\ifx \bissue  \undefined \def \bissue#1{#1}\fi
\ifx \bfpage  \undefined \def \bfpage#1{#1}\fi
\ifx \blpage  \undefined \def \blpage #1{#1}\fi
\ifx \burl  \undefined \def \burl#1{\textsf{#1}}\fi
\ifx \doiurl  \undefined \def \doiurl#1{\url{https://doi.org/#1}}\fi
\ifx \betal  \undefined \def \betal{\textit{et al.}}\fi
\ifx \binstitute  \undefined \def \binstitute#1{#1}\fi
\ifx \binstitutionaled  \undefined \def \binstitutionaled#1{#1}\fi
\ifx \bctitle  \undefined \def \bctitle#1{#1}\fi
\ifx \beditor  \undefined \def \beditor#1{#1}\fi
\ifx \bpublisher  \undefined \def \bpublisher#1{#1}\fi
\ifx \bbtitle  \undefined \def \bbtitle#1{#1}\fi
\ifx \bedition  \undefined \def \bedition#1{#1}\fi
\ifx \bseriesno  \undefined \def \bseriesno#1{#1}\fi
\ifx \blocation  \undefined \def \blocation#1{#1}\fi
\ifx \bsertitle  \undefined \def \bsertitle#1{#1}\fi
\ifx \bsnm \undefined \def \bsnm#1{#1}\fi
\ifx \bsuffix \undefined \def \bsuffix#1{#1}\fi
\ifx \bparticle \undefined \def \bparticle#1{#1}\fi
\ifx \barticle \undefined \def \barticle#1{#1}\fi
\bibcommenthead
\ifx \bconfdate \undefined \def \bconfdate #1{#1}\fi
\ifx \botherref \undefined \def \botherref #1{#1}\fi
\ifx \url \undefined \def \url#1{\textsf{#1}}\fi
\ifx \bchapter \undefined \def \bchapter#1{#1}\fi
\ifx \bbook \undefined \def \bbook#1{#1}\fi
\ifx \bcomment \undefined \def \bcomment#1{#1}\fi
\ifx \oauthor \undefined \def \oauthor#1{#1}\fi
\ifx \citeauthoryear \undefined \def \citeauthoryear#1{#1}\fi
\ifx \endbibitem  \undefined \def \endbibitem {}\fi
\ifx \bconflocation  \undefined \def \bconflocation#1{#1}\fi
\ifx \arxivurl  \undefined \def \arxivurl#1{\textsf{#1}}\fi
\csname PreBibitemsHook\endcsname

\bibitem[\protect\citeauthoryear{J{\o}rgensen et~al.}{2022}]{Jørgensen2022}
\begin{barticle}
\bauthor{\bsnm{J{\o}rgensen}, \binits{A.A.}},
\bauthor{\bsnm{Kong}, \binits{D.}},
\bauthor{\bsnm{Henriksen}, \binits{M.R.}},
\bauthor{\bsnm{Klejs}, \binits{F.}},
\bauthor{\bsnm{Ye}, \binits{Z.}},
\bauthor{\bsnm{Helgason}, \binits{{\`O}.B.}},
\bauthor{\bsnm{Hansen}, \binits{H.E.}},
\bauthor{\bsnm{Hu}, \binits{H.}},
\bauthor{\bsnm{Yankov}, \binits{M.}},
\bauthor{\bsnm{Forchhammer}, \binits{S.}},
\bauthor{\bsnm{Andrekson}, \binits{P.}},
\bauthor{\bsnm{Larsson}, \binits{A.}},
\bauthor{\bsnm{Karlsson}, \binits{M.}},
\bauthor{\bsnm{Schr{\"o}der}, \binits{J.}},
\bauthor{\bsnm{Sasaki}, \binits{Y.}},
\bauthor{\bsnm{Aikawa}, \binits{K.}},
\bauthor{\bsnm{Thomsen}, \binits{J.W.}},
\bauthor{\bsnm{Morioka}, \binits{T.}},
\bauthor{\bsnm{Galili}, \binits{M.}},
\bauthor{\bsnm{{Torres-Company}}, \binits{V.}},
\bauthor{\bsnm{Oxenl{\o}we}, \binits{L.K.}}:
\batitle{Petabit-per-second data transmission using a chip-scale microcomb ring
  resonator source}.
\bjtitle{Nature Photonics}
\bvolume{16}(\bissue{11}),
\bfpage{798}--\blpage{802}
(\byear{2022})
\doiurl{10.1038/s41566-022-01082-z}
\end{barticle}
\endbibitem

\bibitem[\protect\citeauthoryear{Hummon et~al.}{2018}]{Hummon:18}
\begin{barticle}
\bauthor{\bsnm{Hummon}, \binits{M.T.}},
\bauthor{\bsnm{Kang}, \binits{S.}},
\bauthor{\bsnm{Bopp}, \binits{D.G.}},
\bauthor{\bsnm{Li}, \binits{Q.}},
\bauthor{\bsnm{Westly}, \binits{D.A.}},
\bauthor{\bsnm{Kim}, \binits{S.}},
\bauthor{\bsnm{Fredrick}, \binits{C.}},
\bauthor{\bsnm{Diddams}, \binits{S.A.}},
\bauthor{\bsnm{Srinivasan}, \binits{K.}},
\bauthor{\bsnm{Aksyuk}, \binits{V.}},
\bauthor{\bsnm{Kitching}, \binits{J.E.}}:
\batitle{Photonic chip for laser stabilization to an atomic vapor with
  10\&\#x2212;11 instability}.
\bjtitle{Optica}
\bvolume{5}(\bissue{4}),
\bfpage{443}--\blpage{449}
(\byear{2018})
\doiurl{10.1364/OPTICA.5.000443}
\end{barticle}
\endbibitem

\bibitem[\protect\citeauthoryear{Kirchmair et~al.}{2013}]{Kirchmair_2013}
\begin{barticle}
\bauthor{\bsnm{Kirchmair}, \binits{G.}},
\bauthor{\bsnm{Vlastakis}, \binits{B.}},
\bauthor{\bsnm{Leghtas}, \binits{Z.}},
\bauthor{\bsnm{Nigg}, \binits{S.E.}},
\bauthor{\bsnm{Paik}, \binits{H.}},
\bauthor{\bsnm{Ginossar}, \binits{E.}},
\bauthor{\bsnm{Mirrahimi}, \binits{M.}},
\bauthor{\bsnm{Frunzio}, \binits{L.}},
\bauthor{\bsnm{Girvin}, \binits{S.M.}},
\bauthor{\bsnm{Schoelkopf}, \binits{R.J.}}:
\batitle{Observation of quantum state collapse and revival due to the
  single-photon kerr effect}.
\bjtitle{Nature}
\bvolume{495}(\bissue{7440}),
\bfpage{205}--\blpage{209}
(\byear{2013})
\doiurl{10.1038/nature11902}
\end{barticle}
\endbibitem

\bibitem[\protect\citeauthoryear{Wu et~al.}{2024}]{Wu_2024}
\begin{barticle}
\bauthor{\bsnm{Wu}, \binits{T.-H.}},
\bauthor{\bsnm{Ledezma}, \binits{L.}},
\bauthor{\bsnm{Fredrick}, \binits{C.}},
\bauthor{\bsnm{Sekhar}, \binits{P.}},
\bauthor{\bsnm{Sekine}, \binits{R.}},
\bauthor{\bsnm{Guo}, \binits{Q.}},
\bauthor{\bsnm{Briggs}, \binits{R.M.}},
\bauthor{\bsnm{Marandi}, \binits{A.}},
\bauthor{\bsnm{Diddams}, \binits{S.A.}}:
\batitle{Visible-to-ultraviolet frequency comb generation in lithium niobate
  nanophotonic waveguides}.
\bjtitle{Nature Photonics}
\bvolume{18}(\bissue{3}),
\bfpage{218}--\blpage{223}
(\byear{2024})
\doiurl{10.1038/s41566-023-01364-0}
\end{barticle}
\endbibitem

\bibitem[\protect\citeauthoryear{Jankowski et~al.}{2020}]{Jankowski:20}
\begin{barticle}
\bauthor{\bsnm{Jankowski}, \binits{M.}},
\bauthor{\bsnm{Langrock}, \binits{C.}},
\bauthor{\bsnm{Desiatov}, \binits{B.}},
\bauthor{\bsnm{Marandi}, \binits{A.}},
\bauthor{\bsnm{Wang}, \binits{C.}},
\bauthor{\bsnm{Zhang}, \binits{M.}},
\bauthor{\bsnm{Phillips}, \binits{C.R.}},
\bauthor{\bsnm{Lon\v{c}ar}, \binits{M.}},
\bauthor{\bsnm{Fejer}, \binits{M.M.}}:
\batitle{Ultrabroadband nonlinear optics in nanophotonic periodically poled
  lithium niobate waveguides}.
\bjtitle{Optica}
\bvolume{7}(\bissue{1}),
\bfpage{40}--\blpage{46}
(\byear{2020})
\doiurl{10.1364/OPTICA.7.000040}
\end{barticle}
\endbibitem

\bibitem[\protect\citeauthoryear{Carlson et~al.}{2019}]{Carlson:2019}
\begin{barticle}
\bauthor{\bsnm{Carlson}, \binits{D.R.}},
\bauthor{\bsnm{Hutchison}, \binits{P.}},
\bauthor{\bsnm{Hickstein}, \binits{D.D.}},
\bauthor{\bsnm{Papp}, \binits{S.B.}}:
\batitle{Generating few-cycle pulses with integrated nonlinear photonics}.
\bjtitle{Opt. Express}
\bvolume{27}(\bissue{26}),
\bfpage{37374}--\blpage{37382}
(\byear{2019})
\doiurl{10.1364/OE.27.037374}
\end{barticle}
\endbibitem

\bibitem[\protect\citeauthoryear{Leefmans et~al.}{2024}]{Leefmans_2024}
\begin{barticle}
\bauthor{\bsnm{Leefmans}, \binits{C.R.}},
\bauthor{\bsnm{Parto}, \binits{M.}},
\bauthor{\bsnm{Williams}, \binits{J.}},
\bauthor{\bsnm{Li}, \binits{G.H.Y.}},
\bauthor{\bsnm{Dutt}, \binits{A.}},
\bauthor{\bsnm{Nori}, \binits{F.}},
\bauthor{\bsnm{Marandi}, \binits{A.}}:
\batitle{Topological temporally mode-locked laser}.
\bjtitle{Nature Physics}
(\byear{2024})
\doiurl{10.1038/s41567-024-02420-4}
\end{barticle}
\endbibitem

\bibitem[\protect\citeauthoryear{Briles et~al.}{2018}]{Briles2018}
\begin{barticle}
\bauthor{\bsnm{Briles}, \binits{T.C.}},
\bauthor{\bsnm{Stone}, \binits{J.R.}},
\bauthor{\bsnm{Drake}, \binits{T.E.}},
\bauthor{\bsnm{Spencer}, \binits{D.T.}},
\bauthor{\bsnm{Fredrick}, \binits{C.}},
\bauthor{\bsnm{Li}, \binits{Q.}},
\bauthor{\bsnm{Westly}, \binits{D.}},
\bauthor{\bsnm{Ilic}, \binits{B.R.}},
\bauthor{\bsnm{Srinivasan}, \binits{K.}},
\bauthor{\bsnm{Diddams}, \binits{S.A.}},
\bauthor{\bsnm{Papp}, \binits{S.B.}}:
\batitle{Interlocking {{Kerr-microresonator}} frequency combs for microwave to
  optical synthesis}.
\bjtitle{Optics Letters}
\bvolume{43}(\bissue{12}),
\bfpage{2933}--\blpage{2936}
(\byear{2018})
\doiurl{10.1364/OL.43.002933}
\end{barticle}
\endbibitem

\bibitem[\protect\citeauthoryear{Zang et~al.}{2024}]{zang2024laserpower}
\begin{botherref}
\oauthor{\bsnm{Zang}, \binits{J.}},
\oauthor{\bsnm{Yu}, \binits{S.-P.}},
\oauthor{\bsnm{Liu}, \binits{H.}},
\oauthor{\bsnm{Jin}, \binits{Y.}},
\oauthor{\bsnm{Briles}, \binits{T.C.}},
\oauthor{\bsnm{Carlson}, \binits{D.R.}},
\oauthor{\bsnm{Papp}, \binits{S.B.}}:
Laser-power consumption of soliton formation in a bidirectional Kerr resonator
(2024).
\url{https://doi.org/10.48550/arXiv.2401.16740}
\end{botherref}
\endbibitem

\bibitem[\protect\citeauthoryear{Black et~al.}{2023}]{Black2023}
\begin{barticle}
\bauthor{\bsnm{Black}, \binits{J.A.}},
\bauthor{\bsnm{Newman}, \binits{Z.L.}},
\bauthor{\bsnm{Yu}, \binits{S.-P.}},
\bauthor{\bsnm{Carlson}, \binits{D.R.}},
\bauthor{\bsnm{Papp}, \binits{S.B.}}:
\batitle{Nonlinear networks for arbitrary optical synthesis}.
\bjtitle{Phys. Rev. X}
\bvolume{13},
\bfpage{021027}
(\byear{2023})
\doiurl{10.1103/PhysRevX.13.021027}
\end{barticle}
\endbibitem

\bibitem[\protect\citeauthoryear{Domeneguetti et~al.}{2021}]{Domeneguetti:21}
\begin{barticle}
\bauthor{\bsnm{Domeneguetti}, \binits{R.R.}},
\bauthor{\bsnm{Zhao}, \binits{Y.}},
\bauthor{\bsnm{Ji}, \binits{X.}},
\bauthor{\bsnm{Martinelli}, \binits{M.}},
\bauthor{\bsnm{Lipson}, \binits{M.}},
\bauthor{\bsnm{Gaeta}, \binits{A.L.}},
\bauthor{\bsnm{Nussenzveig}, \binits{P.}}:
\batitle{Parametric sideband generation in cmos-compatible oscillators from
  visible to telecom wavelengths}.
\bjtitle{Optica}
\bvolume{8}(\bissue{3}),
\bfpage{316}--\blpage{322}
(\byear{2021})
\doiurl{10.1364/OPTICA.404755}
\end{barticle}
\endbibitem

\bibitem[\protect\citeauthoryear{Cai
  et~al.}{2022}]{Cai2022OctavespanningMicrocomb}
\begin{barticle}
\bauthor{\bsnm{Cai}, \binits{L.}},
\bauthor{\bsnm{Li}, \binits{J.}},
\bauthor{\bsnm{Wang}, \binits{R.}},
\bauthor{\bsnm{Li}, \binits{Q.}}:
\batitle{Octave-spanning microcomb generation in
  {{4H-silicon-carbide-on-insulator}} photonics platform}.
\bjtitle{Photonics Research}
\bvolume{10}(\bissue{4}),
\bfpage{870}--\blpage{876}
(\byear{2022})
\doiurl{10.1364/PRJ.449267} .
Accessed 2023-09-11
\end{barticle}
\endbibitem

\bibitem[\protect\citeauthoryear{Rao et~al.}{2021}]{Rao_2021}
\begin{botherref}
\oauthor{\bsnm{Rao}, \binits{A.}},
\oauthor{\bsnm{Moille}, \binits{G.}},
\oauthor{\bsnm{Lu}, \binits{X.}},
\oauthor{\bsnm{Westly}, \binits{D.A.}},
\oauthor{\bsnm{Sacchetto}, \binits{D.}},
\oauthor{\bsnm{Geiselmann}, \binits{M.}},
\oauthor{\bsnm{Zervas}, \binits{M.}},
\oauthor{\bsnm{Papp}, \binits{S.B.}},
\oauthor{\bsnm{Bowers}, \binits{J.}},
\oauthor{\bsnm{Srinivasan}, \binits{K.}}:
Towards integrated photonic interposers for processing octave-spanning
  microresonator frequency combs.
Light: Science \& Applications
\textbf{10}(1)
(2021)
\doiurl{10.1038/s41377-021-00549-y}
\end{botherref}
\endbibitem

\bibitem[\protect\citeauthoryear{Gu et~al.}{2023}]{Gu:23}
\begin{barticle}
\bauthor{\bsnm{Gu}, \binits{J.}},
\bauthor{\bsnm{Li}, \binits{X.}},
\bauthor{\bsnm{Qi}, \binits{K.}},
\bauthor{\bsnm{Pu}, \binits{K.}},
\bauthor{\bsnm{Li}, \binits{Z.}},
\bauthor{\bsnm{Zhang}, \binits{F.}},
\bauthor{\bsnm{Li}, \binits{T.}},
\bauthor{\bsnm{Xie}, \binits{Z.}},
\bauthor{\bsnm{Xiao}, \binits{M.}},
\bauthor{\bsnm{Jiang}, \binits{X.}}:
\batitle{Octave-spanning soliton microcomb in silica microdisk resonators}.
\bjtitle{Opt. Lett.}
\bvolume{48}(\bissue{5}),
\bfpage{1100}--\blpage{1103}
(\byear{2023})
\doiurl{10.1364/OL.479251}
\end{barticle}
\endbibitem

\bibitem[\protect\citeauthoryear{Liu et~al.}{2021}]{Liu_2021}
\begin{botherref}
\oauthor{\bsnm{Liu}, \binits{X.}},
\oauthor{\bsnm{Gong}, \binits{Z.}},
\oauthor{\bsnm{Bruch}, \binits{A.W.}},
\oauthor{\bsnm{Surya}, \binits{J.B.}},
\oauthor{\bsnm{Lu}, \binits{J.}},
\oauthor{\bsnm{Tang}, \binits{H.X.}}:
Aluminum nitride nanophotonics for beyond-octave soliton microcomb generation
  and self-referencing.
Nature Communications
\textbf{12}(1)
(2021)
\doiurl{10.1038/s41467-021-25751-9}
\end{botherref}
\endbibitem

\bibitem[\protect\citeauthoryear{Gong et~al.}{2020}]{Gong:20}
\begin{barticle}
\bauthor{\bsnm{Gong}, \binits{Z.}},
\bauthor{\bsnm{Liu}, \binits{X.}},
\bauthor{\bsnm{Xu}, \binits{Y.}},
\bauthor{\bsnm{Tang}, \binits{H.X.}}:
\batitle{Near-octave lithium niobate soliton microcomb}.
\bjtitle{Optica}
\bvolume{7}(\bissue{10}),
\bfpage{1275}--\blpage{1278}
(\byear{2020})
\doiurl{10.1364/OPTICA.400994}
\end{barticle}
\endbibitem

\bibitem[\protect\citeauthoryear{Lugiato and Lefever}{1987}]{Lugiato1987}
\begin{barticle}
\bauthor{\bsnm{Lugiato}, \binits{L.A.}},
\bauthor{\bsnm{Lefever}, \binits{R.}}:
\batitle{Spatial {{Dissipative Structures}} in {{Passive Optical Systems}}}.
\bjtitle{Physical Review Letters}
\bvolume{58}(\bissue{21}),
\bfpage{2209}--\blpage{2211}
(\byear{1987})
\doiurl{10.1103/PhysRevLett.58.2209} .
Accessed 2014-06-30
\end{barticle}
\endbibitem

\bibitem[\protect\citeauthoryear{Lucas
  et~al.}{2023}]{Lucas2023TailoringMicrocombs}
\begin{barticle}
\bauthor{\bsnm{Lucas}, \binits{E.}},
\bauthor{\bsnm{Yu}, \binits{S.-P.}},
\bauthor{\bsnm{Briles}, \binits{T.C.}},
\bauthor{\bsnm{Carlson}, \binits{D.R.}},
\bauthor{\bsnm{Papp}, \binits{S.B.}}:
\batitle{Tailoring microcombs with inverse-designed, meta-dispersion
  microresonators}.
\bjtitle{Nature Photonics}
\bvolume{17}(\bissue{11}),
\bfpage{943}--\blpage{950}
(\byear{2023})
\doiurl{10.1038/s41566-023-01252-7} .
Accessed 2023-11-02
\end{barticle}
\endbibitem

\bibitem[\protect\citeauthoryear{Okawachi et~al.}{2014}]{Okawachi2014}
\begin{barticle}
\bauthor{\bsnm{Okawachi}, \binits{Y.}},
\bauthor{\bsnm{Lamont}, \binits{M.R.E.}},
\bauthor{\bsnm{Luke}, \binits{K.}},
\bauthor{\bsnm{Carvalho}, \binits{D.O.}},
\bauthor{\bsnm{Yu}, \binits{M.}},
\bauthor{\bsnm{Lipson}, \binits{M.}},
\bauthor{\bsnm{Gaeta}, \binits{A.L.}}:
\batitle{Bandwidth shaping of microresonator-based frequency combs via
  dispersion engineering}.
\bjtitle{Optics Letters}
\bvolume{39}(\bissue{12}),
\bfpage{3535}--\blpage{3538}
(\byear{2014})
\doiurl{10.1364/OL.39.003535}
\end{barticle}
\endbibitem

\bibitem[\protect\citeauthoryear{Lucas et~al.}{2017}]{Lucas2017}
\begin{barticle}
\bauthor{\bsnm{Lucas}, \binits{E.}},
\bauthor{\bsnm{Guo}, \binits{H.}},
\bauthor{\bsnm{Jost}, \binits{J.D.}},
\bauthor{\bsnm{Karpov}, \binits{M.}},
\bauthor{\bsnm{Kippenberg}, \binits{T.J.}}:
\batitle{Detuning-dependent properties and dispersion-induced instabilities of
  temporal dissipative {{Kerr}} solitons in optical microresonators}.
\bjtitle{Physical Review A}
\bvolume{95}(\bissue{4}),
\bfpage{043822}
(\byear{2017})
\doiurl{10.1103/PhysRevA.95.043822} .
Accessed 2017-03-31
\end{barticle}
\endbibitem

\bibitem[\protect\citeauthoryear{Cole et~al.}{2017}]{Cole2016a}
\begin{barticle}
\bauthor{\bsnm{Cole}, \binits{D.C.}},
\bauthor{\bsnm{Lamb}, \binits{E.S.}},
\bauthor{\bsnm{Del'Haye}, \binits{P.}},
\bauthor{\bsnm{Diddams}, \binits{S.A.}},
\bauthor{\bsnm{Papp}, \binits{S.B.}}:
\batitle{Soliton crystals in {{Kerr}} resonators}.
\bjtitle{Nature Photonics}
\bvolume{11}(\bissue{10}),
\bfpage{671}--\blpage{676}
(\byear{2017})
\doiurl{10.1038/s41566-017-0009-z}
\end{barticle}
\endbibitem

\bibitem[\protect\citeauthoryear{Yuan et~al.}{2023}]{Yuan2023SolitonPulse}
\begin{botherref}
\oauthor{\bsnm{Yuan}, \binits{Z.}},
\oauthor{\bsnm{Gao}, \binits{M.}},
\oauthor{\bsnm{Yu}, \binits{Y.}},
\oauthor{\bsnm{Wang}, \binits{H.}},
\oauthor{\bsnm{Jin}, \binits{W.}},
\oauthor{\bsnm{Ji}, \binits{Q.-X.}},
\oauthor{\bsnm{Feshali}, \binits{A.}},
\oauthor{\bsnm{Paniccia}, \binits{M.}},
\oauthor{\bsnm{Bowers}, \binits{J.}},
\oauthor{\bsnm{Vahala}, \binits{K.}}:
Soliton pulse pairs at multiple colours in normal dispersion microresonators.
Nature Photonics,
1--7
(2023)
\doiurl{10.1038/s41566-023-01257-2} .
Accessed 2023-08-05
\end{botherref}
\endbibitem

\bibitem[\protect\citeauthoryear{Wan et~al.}{2023}]{Wan_2023}
\begin{botherref}
\oauthor{\bsnm{Wan}, \binits{S.}},
\oauthor{\bsnm{Wang}, \binits{P.}},
\oauthor{\bsnm{Ma}, \binits{R.}},
\oauthor{\bsnm{Wang}, \binits{Z.}},
\oauthor{\bsnm{Niu}, \binits{R.}},
\oauthor{\bsnm{He}, \binits{D.}},
\oauthor{\bsnm{Guo}, \binits{G.}},
\oauthor{\bsnm{Bo}, \binits{F.}},
\oauthor{\bsnm{Liu}, \binits{J.}},
\oauthor{\bsnm{Dong}, \binits{C.}}:
Photorefraction‐assisted self‐emergence of dissipative kerr solitons.
Laser \& Photonics Reviews
\textbf{18}(2)
(2023)
\doiurl{10.1002/lpor.202300627}
\end{botherref}
\endbibitem

\bibitem[\protect\citeauthoryear{Xu et~al.}{2021}]{Xu:21}
\begin{barticle}
\bauthor{\bsnm{Xu}, \binits{Y.}},
\bauthor{\bsnm{Shen}, \binits{M.}},
\bauthor{\bsnm{Lu}, \binits{J.}},
\bauthor{\bsnm{Surya}, \binits{J.B.}},
\bauthor{\bsnm{Sayem}, \binits{A.A.}},
\bauthor{\bsnm{Tang}, \binits{H.X.}}:
\batitle{Mitigating photorefractive effect in thin-film lithium niobate
  microring resonators}.
\bjtitle{Opt. Express}
\bvolume{29}(\bissue{4}),
\bfpage{5497}--\blpage{5504}
(\byear{2021})
\doiurl{10.1364/OE.418877}
\end{barticle}
\endbibitem

\bibitem[\protect\citeauthoryear{Yu et~al.}{2021}]{Yu2020}
\begin{barticle}
\bauthor{\bsnm{Yu}, \binits{S.-P.}},
\bauthor{\bsnm{Cole}, \binits{D.C.}},
\bauthor{\bsnm{Jung}, \binits{H.}},
\bauthor{\bsnm{Moille}, \binits{G.T.}},
\bauthor{\bsnm{Srinivasan}, \binits{K.}},
\bauthor{\bsnm{Papp}, \binits{S.B.}}:
\batitle{Spontaneous pulse formation in edgeless photonic crystal resonators}.
\bjtitle{Nature Photonics}
\bvolume{15}(\bissue{6}),
\bfpage{461}--\blpage{467}
(\byear{2021})
\doiurl{10.1038/s41566-021-00800-3}
\end{barticle}
\endbibitem

\bibitem[\protect\citeauthoryear{Black et~al.}{2022}]{Black2022}
\begin{barticle}
\bauthor{\bsnm{Black}, \binits{J.A.}},
\bauthor{\bsnm{Brodnik}, \binits{G.}},
\bauthor{\bsnm{Liu}, \binits{H.}},
\bauthor{\bsnm{Yu}, \binits{S.-P.}},
\bauthor{\bsnm{Carlson}, \binits{D.R.}},
\bauthor{\bsnm{Zang}, \binits{J.}},
\bauthor{\bsnm{Briles}, \binits{T.C.}},
\bauthor{\bsnm{Papp}, \binits{S.B.}}:
\batitle{Optical-parametric oscillation in photonic-crystal ring resonators}.
\bjtitle{Optica}
\bvolume{9}(\bissue{10}),
\bfpage{1183}
(\byear{2022})
\doiurl{10.1364/OPTICA.469210}
\end{barticle}
\endbibitem

\bibitem[\protect\citeauthoryear{Liu et~al.}{2024}]{Liu2023Threshold}
\begin{barticle}
\bauthor{\bsnm{Liu}, \binits{H.}},
\bauthor{\bsnm{Brodnik}, \binits{G.M.}},
\bauthor{\bsnm{Zang}, \binits{J.}},
\bauthor{\bsnm{Carlson}, \binits{D.R.}},
\bauthor{\bsnm{Black}, \binits{J.A.}},
\bauthor{\bsnm{Papp}, \binits{S.B.}}:
\batitle{Threshold and laser conversion in nanostructured-resonator parametric
  oscillators}.
\bjtitle{Phys. Rev. Lett.}
\bvolume{132},
\bfpage{023801}
(\byear{2024})
\doiurl{10.1103/PhysRevLett.132.023801}
\end{barticle}
\endbibitem

\bibitem[\protect\citeauthoryear{Yu et~al.}{2022}]{Yu2021}
\begin{barticle}
\bauthor{\bsnm{Yu}, \binits{S.-P.}},
\bauthor{\bsnm{Lucas}, \binits{E.}},
\bauthor{\bsnm{Zang}, \binits{J.}},
\bauthor{\bsnm{Papp}, \binits{S.B.}}:
\batitle{A continuum of bright and dark-pulse states in a photonic-crystal
  resonator}.
\bjtitle{Nature Communications}
\bvolume{13}(\bissue{1}),
\bfpage{3134}
(\byear{2022})
\doiurl{10.1038/s41467-022-30774-x}
\end{barticle}
\endbibitem

\bibitem[\protect\citeauthoryear{Stone
  et~al.}{2023}]{Stone2023WavelengthaccurateNonlinear}
\begin{botherref}
\oauthor{\bsnm{Stone}, \binits{J.R.}},
\oauthor{\bsnm{Lu}, \binits{X.}},
\oauthor{\bsnm{Moille}, \binits{G.}},
\oauthor{\bsnm{Westly}, \binits{D.}},
\oauthor{\bsnm{Rahman}, \binits{T.}},
\oauthor{\bsnm{Srinivasan}, \binits{K.}}:
Wavelength-accurate nonlinear conversion through wavenumber selectivity in
  photonic crystal resonators.
Nature Photonics,
1--8
(2023)
\doiurl{10.1038/s41566-023-01326-6} .
Accessed 2023-11-20
\end{botherref}
\endbibitem

\bibitem[\protect\citeauthoryear{Li
  et~al.}{2023}]{Li2023SymmetricallyDispersionengineered}
\begin{barticle}
\bauthor{\bsnm{Li}, \binits{J.}},
\bauthor{\bsnm{Zhang}, \binits{Y.}},
\bauthor{\bsnm{Xie}, \binits{Y.}},
\bauthor{\bsnm{Lin}, \binits{S.}},
\bauthor{\bsnm{Zeng}, \binits{S.}},
\bauthor{\bsnm{Wu}, \binits{Z.}},
\bauthor{\bsnm{Yu}, \binits{S.}}:
\batitle{Symmetrically dispersion-engineered microcombs}.
\bjtitle{Communications Physics}
\bvolume{6}(\bissue{1}),
\bfpage{1}--\blpage{9}
(\byear{2023})
\doiurl{10.1038/s42005-023-01453-0} .
Accessed 2023-11-23
\end{barticle}
\endbibitem

\bibitem[\protect\citeauthoryear{Helgason
  et~al.}{2023}]{Helgason2023SurpassingNonlinear}
\begin{botherref}
\oauthor{\bsnm{Helgason}, \binits{{\'O}.B.}},
\oauthor{\bsnm{Girardi}, \binits{M.}},
\oauthor{\bsnm{Ye}, \binits{Z.}},
\oauthor{\bsnm{Lei}, \binits{F.}},
\oauthor{\bsnm{Schr{\"o}der}, \binits{J.}},
\oauthor{\bsnm{{Torres-Company}}, \binits{V.}}:
Surpassing the nonlinear conversion efficiency of soliton microcombs.
Nature Photonics,
1--8
(2023)
\doiurl{10.1038/s41566-023-01280-3} .
Accessed 2023-09-04
\end{botherref}
\endbibitem

\bibitem[\protect\citeauthoryear{{Rebolledo-Salgado}
  et~al.}{2023}]{Rebolledo-Salgado2023PlaticonDynamics}
\begin{barticle}
\bauthor{\bsnm{{Rebolledo-Salgado}}, \binits{I.}},
\bauthor{\bsnm{{Quevedo-Gal{\'a}n}}, \binits{C.}},
\bauthor{\bsnm{Helgason}, \binits{{\'O}.B.}},
\bauthor{\bsnm{L{\"o}{\"o}f}, \binits{A.}},
\bauthor{\bsnm{Ye}, \binits{Z.}},
\bauthor{\bsnm{Lei}, \binits{F.}},
\bauthor{\bsnm{Schr{\"o}der}, \binits{J.}},
\bauthor{\bsnm{Zelan}, \binits{M.}},
\bauthor{\bsnm{{Torres-Company}}, \binits{V.}}:
\batitle{Platicon dynamics in photonic molecules}.
\bjtitle{Communications Physics}
\bvolume{6}(\bissue{1}),
\bfpage{303}
(\byear{2023})
\doiurl{10.1038/s42005-023-01424-5} .
Accessed 2023-10-27
\end{barticle}
\endbibitem

\bibitem[\protect\citeauthoryear{Kondratiev and
  Lobanov}{2020}]{Kondratiev2020ModulationalInstability}
\begin{barticle}
\bauthor{\bsnm{Kondratiev}, \binits{N.M.}},
\bauthor{\bsnm{Lobanov}, \binits{V.E.}}:
\batitle{Modulational instability and frequency combs in
  whispering-gallery-mode microresonators with backscattering}.
\bjtitle{Physical Review A}
\bvolume{101}(\bissue{1}),
\bfpage{013816}
(\byear{2020})
\doiurl{10.1103/PhysRevA.101.013816} .
Accessed 2023-05-23
\end{barticle}
\endbibitem

\bibitem[\protect\citeauthoryear{Jung et~al.}{2019}]{Jung2019}
\begin{bchapter}
\bauthor{\bsnm{Jung}, \binits{H.}},
\bauthor{\bsnm{Yu}, \binits{S.-P.}},
\bauthor{\bsnm{Carlson}, \binits{D.R.}},
\bauthor{\bsnm{Drake}, \binits{T.E.}},
\bauthor{\bsnm{Briles}, \binits{T.C.}},
\bauthor{\bsnm{Papp}, \binits{S.B.}}:
\bctitle{Kerr {{Solitons}} with {{Tantala Ring Resonators}}}.
In: \bbtitle{Nonlinear {{Optics}} ({{NLO}})},
pp. \bfpage{2}--\blpage{3}.
\bpublisher{{OSA}},
\blocation{{Washington, D.C., D.C.}}
(\byear{2019}).
\doiurl{10.1364/NLO.2019.NW2A.3} .
\burl{https://www.osapublishing.org/abstract.cfm?URI=NLO-2019-NW2A.3}
\end{bchapter}
\endbibitem

\bibitem[\protect\citeauthoryear{Black et~al.}{2021}]{Black2021}
\begin{barticle}
\bauthor{\bsnm{Black}, \binits{J.A.}},
\bauthor{\bsnm{Streater}, \binits{R.}},
\bauthor{\bsnm{Lamee}, \binits{K.F.}},
\bauthor{\bsnm{Carlson}, \binits{D.R.}},
\bauthor{\bsnm{Yu}, \binits{S.-P.}},
\bauthor{\bsnm{Papp}, \binits{S.B.}}:
\batitle{Group-velocity-dispersion engineering of tantala integrated
  photonics}.
\bjtitle{Optics Letters}
\bvolume{46}(\bissue{4}),
\bfpage{817}
(\byear{2021})
\doiurl{10.1364/OL.414095}
\end{barticle}
\endbibitem

\bibitem[\protect\citeauthoryear{Lu et~al.}{2020}]{Lu2020}
\begin{barticle}
\bauthor{\bsnm{Lu}, \binits{X.}},
\bauthor{\bsnm{Rao}, \binits{A.}},
\bauthor{\bsnm{Moille}, \binits{G.}},
\bauthor{\bsnm{Westly}, \binits{D.A.}},
\bauthor{\bsnm{Srinivasan}, \binits{K.}}:
\batitle{Universal frequency engineering tool for microcavity nonlinear optics:
  Multiple selective mode splitting of whispering-gallery resonances}.
\bjtitle{Photonics Research}
\bvolume{8}(\bissue{11}),
\bfpage{1676}
(\byear{2020})
\doiurl{10.1364/PRJ.401755}
\end{barticle}
\endbibitem

\bibitem[\protect\citeauthoryear{Garbin et~al.}{2017}]{Garbin2017}
\begin{barticle}
\bauthor{\bsnm{Garbin}, \binits{B.}},
\bauthor{\bsnm{Wang}, \binits{Y.}},
\bauthor{\bsnm{Murdoch}, \binits{S.G.}},
\bauthor{\bsnm{Oppo}, \binits{G.-L.}},
\bauthor{\bsnm{Coen}, \binits{S.}},
\bauthor{\bsnm{Erkintalo}, \binits{M.}}:
\batitle{Experimental and numerical investigations of switching wave dynamics
  in a normally dispersive fibre ring resonator}.
\bjtitle{The European Physical Journal D}
\bvolume{71}(\bissue{9}),
\bfpage{240}
(\byear{2017})
\doiurl{10.1140/epjd/e2017-80133-7}
\end{barticle}
\endbibitem

\end{thebibliography}

\end{document}